\documentclass[%
reprint,
superscriptaddress,
showpacs,preprintnumbers,
showkeys,
nofootinbib,
bibnotes,
longbibliography,
amsmath,amssymb,
aps,
prl
]{revtex4-1}
\usepackage{graphicx}
\usepackage{dcolumn}
\usepackage{bm}
\usepackage{isotope}
\usepackage{siunitx}
\usepackage{color}
\usepackage{multirow}
\usepackage{ulem}
\usepackage[colorlinks,breaklinks]{hyperref}
\AtBeginDocument{%
  \hypersetup{
    citecolor=blue,
    linkcolor=blue,   
    urlcolor=blue}}

\definecolor{red4}{rgb}{0.57917,0.,0.}
\definecolor{green4}{rgb}{0.,0.57917,0.}
\definecolor{blue4}{rgb}{0.,0.,0.57917}
\definecolor{purple4}{rgb}{0.333333,0.10196,0.545098}

\newcommand{\strikethrough}[1]{\textcolor{red4}{\sout{#1}}}

\renewcommand{\strikethrough}[1]{\ignorespaces}

\newcommand{\geant}{\textsc{Geant4}}
\newcommand{\triumf}{{\scshape Triumf}}
\newcommand{\isac}{{\scshape Isac}}
\newcommand{\trinat}{{\scshape Trinat}}

\newcommand{\tsups}[1]{\textsuperscript{#1}}
\newcommand{\zerotozero}{\mbox{$0^+\!\!\rightarrow 0^+$}}
\newcommand{\opara}[4]{\begin{picture}(0,0)%
    \put(#1,#2){\makebox[0mm]{%
        \rotatebox{#3}{\smash{%
            $\left.\mbox{\rule{0mm}{#4}}\right\{$} }}}%
  \end{picture} }

\renewcommand{\vec}[1]{\boldsymbol{#1}}

\graphicspath{{.}{FIGS/}}

\AtBeginDocument{\usepackage{booktabs}}

\begin{document}

\title{Precision Measurement of the \boldmath$\beta$ Asymmetry in Spin-Polarized 
  \isotope[37]{K} Decay}

\author{B.~Fenker}
\affiliation{Cyclotron Institute, Texas A\&M University, 3366 TAMU, 
  College Station, Texas 77843-3366, USA}
\affiliation{Department of Physics and Astronomy, Texas A\&M University, 
  4242 TAMU, College Station, Texas 77843-4242, USA}
\author{A.~Gorelov}
\affiliation{TRIUMF, 4004 Wesbrook Mall, Vancouver, British Columbia 
   V6T 2A3, Canada}
\author{D.~Melconian}
\email{dmelconian@tamu.edu}
\affiliation{Cyclotron Institute, Texas A\&M University, 3366 TAMU, 
  College Station, Texas 77843-3366, USA}
\affiliation{Department of Physics and Astronomy, Texas A\&M University, 
  4242 TAMU, College Station, Texas 77843-4242, USA}
\author{J.A.~Behr}
\affiliation{TRIUMF, 4004 Wesbrook Mall, Vancouver, British Columbia V6T 2A3, 
  Canada}
\author{M. Anholm}
\affiliation{TRIUMF, 4004 Wesbrook Mall, Vancouver, British Columbia  V6T 2A3, 
  Canada}
\affiliation{Department of Physics and Astronomy, University of Manitoba, 
  Winnipeg, Manitoba R3T 2N2, Canada}
\author{D.~Ashery} 
\affiliation{School of Physics and Astronomy, Tel Aviv University, 
  Tel Aviv 69978, Israel}
\author{R.S.~Behling}
\affiliation{Cyclotron Institute, Texas A\&M University, 3366 TAMU, 
  College Station, Texas 77843-3366, USA}
\affiliation{Department of Chemistry, Texas A\&M University, 3012 TAMU, 
  College Station, Texas 77843-3012, USA}
\author{I.~Cohen} 
\affiliation{School of Physics and Astronomy, Tel Aviv University, 
  Tel Aviv 69978, Israel}
\author{I.~Craiciu}
\affiliation{TRIUMF, 4004 Wesbrook Mall, Vancouver, British Columbia V6T 2A3, 
  Canada}
\author{G.~Gwinner}
\affiliation{Department of Physics and Astronomy, University of Manitoba, 
  Winnipeg, Manitoba R3T 2N2, Canada}
\author{J.~McNeil}
\affiliation{Department of Physics and Astronomy, University of British 
  Columbia, Vancouver, British Columbia V6T 1Z1, Canada} 
\affiliation{TRIUMF, 4004 Wesbrook Mall, Vancouver, British Columbia V6T 2A3, 
  Canada}
\author{M.~Mehlman}
\affiliation{Cyclotron Institute, Texas A\&M University, 3366 TAMU, 
  College Station, Texas 77843-3366, USA}
\affiliation{Department of Physics and Astronomy, Texas A\&M University, 
  4242 TAMU, College Station, Texas 77843-4242, USA}
\author{K.~Olchanski}
\affiliation{TRIUMF, 4004 Wesbrook Mall, Vancouver, British Columbia V6T 2A3, 
  Canada}
\author{P.D.~Shidling}
\affiliation{Cyclotron Institute, Texas A\&M University, 3366 TAMU, 
  College Station, Texas 77843-3366, USA}
\author{S.~Smale}
\affiliation{TRIUMF, 4004 Wesbrook Mall, Vancouver, British Columbia V6T 2A3, 
  Canada}
\author{C.L.~Warner}
\affiliation{TRIUMF, 4004 Wesbrook Mall, Vancouver, British Columbia V6T 2A3, 
  Canada}

\date{\today}

\begin{abstract}
Using \triumf's neutral atom trap, \trinat, for nuclear $\beta$ decay, 
we have measured the $\beta$ asymmetry with respect to the initial nuclear 
spin in \isotope[37]{K} to be 
$A_\beta = -0.5707\left(13\right)_{\mathrm{syst}}\left(13\right)_{\mathrm{stat}}
\left(5\right)_{\mathrm{pol}}$, a 0.3\% measurement.  
This is the best relative accuracy of any $\beta$-asymmetry measurement in 
a nucleus or the neutron, and is in agreement with the standard model 
prediction $-0.5706(7)$.  We compare constraints on physics beyond the 
standard model with other $\beta$-decay measurements, and improve the value 
of $V_\mathrm{ud}$ measured in this mirror nucleus by a factor of 4.

\end{abstract}

\maketitle

Nuclear $\beta$-decay correlation experiments were instrumental in establishing 
the standard model (SM) charged weak interaction as a theory with spin-$1$ 
$W^\pm$ bosons, coupling only to left-handed neutrinos through a vector 
minus axial-vector ($V\!-\!A$) current. Precision measurements continue 
to probe this structure~\cite{Holstein2014}.  Extensions to the SM propose 
that parity symmetry, which is maximally violated in the weak interaction, 
is restored at some higher energy scale by extending the 
$SU(2)_L\otimes U(1)_Y$ electroweak gauge group to include a right-handed 
$SU(2)_R$ sector.  Manifest left-right symmetric models have an 
angle $\zeta$ which mixes the weak ($W_{L,R}$) eigenstates to form mass 
eigenstates with masses $M_{1,2}$, characterized by 
$\delta=(M_1/M_2)^2$~\cite{Thomas2001}.

Atom and ion trapping techniques~\cite{Behr2014,Sternberg2015,Mehlman2015,
  Ban2013}, and progress in neutron decay 
measurements~\cite{Mund2013,Young2014}, have allowed correlation parameters 
in $\beta$ decay to be measured with improved precision \strikethrough{in 
recent years} recently, 
increasing their sensitivity as probes of non-SM physics.  We present here 
an experiment combining a magneto-optical trap (MOT) with optical pumping 
(OP) to produce a set of nearly ideal conditions: an 
isomerically selected source of highly polarized~\cite{Fenker-NJP-2016} 
$\beta$-decaying atoms that are cold and localized \strikethrough{in} 
within an exceptionally open 
geometry.  We measure the correlation between the spin of a parent 
\isotope[37]{K} nucleus and the momentum of the outgoing $\beta^+$, 
given by the decay rate~\cite{Jackson1957a,*Jackson1957b}:
\begin{equation}
  \frac{d^3\Gamma_{\mathrm{angular}}}{dE_\beta d\Omega_\beta} \propto 
  1+b\frac{m_e}{E_\beta}+\vec{P}_\mathrm{nucl}\!\cdot\!\Big(A_\beta
  \frac{\vec{p}_\beta}{E_\beta}\Big),\label{eq:kin}
\end{equation}
where we have neglected terms that cancel in the asymmetry measurement 
of our geometry.  In this expression, $m_e$, $E_\beta$, and 
$\vec{p}_\beta$ are the mass, 
total energy, and momentum of the positron, $\vec{P}_\mathrm{nucl}$ is the 
polarization of the parent nucleus, and $b$ and $A_\beta$ are correlation 
parameters whose values depend on the symmetries inherent in the weak 
interaction.  We take the SM value $b=0$ for this \strikethrough{letter} 
Letter, consistent with the $E_\beta$ dependence of our observed 
asymmetry as shown below.  We will consider non-SM physics that depend 
on $E_\beta$ in a future publication~\cite{Fenker2017b}.  

The $\beta$ asymmetry has been measured previously in the neutron and ten 
different nuclei.  The focus of this work is the mixed 
$I^\pi=3/2^+\!\rightarrow3/2^+$ Fermi/Gamow-Teller $\beta^+$ decay of 
\isotope[37]{K}, which has a half-life of 
\SI{1.236\,51(94)}{\second}~\cite{Shidling2014} and 
$Q_\mathrm{EC}=\SI{6.147\,47(23)}{\mega\electronvolt}$~\cite{AME2016}.  The 
transition to the ground state of \isotope[37]{Ar} dominates with a 
branching ratio of 97.99(14)\%~\cite{Severijns2008}.  The next most 
significant branch is to an excited $5/2^+$ state at 
\SI{2.7961}{\mega\electronvolt}, which must be pure GT with a value of 
$A_\beta^\mathrm{GT}=-0.6$.  All other branches to excited states are below 
$0.03\%$~\cite{Hagberg1997}.

The corrected comparative half-life for \isotope[37]{K} is 
$\mathcal{F}t=4605.4\pm8.2~\mathrm{s}$~\cite{Shidling2014} based on the 
half-life, branching ratio and $Q_\mathrm{EC}$ values given above.  The 
$\mathcal{F}t$ values for transitions between $T=1/2$ isospin doublets in 
mirror nuclei are related to the $\mathcal{F}t$ value for \zerotozero{} 
decays via:
\begin{equation}
  \mathcal{F}t^\mathrm{mirror}=\frac{2\mathcal{F}t^{\,0^+\!\rightarrow0^+}}{1+
    \frac{f_A}{f_V}\rho^2},\label{eq:Ft-ratio}
\end{equation}
where $f_A/f_V=1.0046(9)$~\cite{Severijns2008} is the ratio of statistical 
rate functions for axial-vector and vector currents, and $\rho=\frac{C_AM_{GT}}{C_VM_F}$ 
is the ratio of Gamow-Teller and Fermi coupling constants ($C_A/C_V$) and 
matrix elements ($M_{GT}/M_F$).  Equation~\eqref{eq:Ft-ratio} with 
$\mathcal{F}t^{\,0^+\!\rightarrow0^+}\!=3072.27(72)~\mathrm{s}$~\cite{Hardy2015} 
leads to $\rho=0.5768(21)$.

\strikethrough{%
 In addition to depending on $\rho$ and the initial nuclear spin, $I$, the 
 values of the correlation parameters for $\beta$ decay depend on the 
 symmetry structure of the underlying interaction, and thereby physics beyond 
 the SM.}
For mixed transitions, the $\beta$ asymmetry including the 
possibility of right-handed currents is given 
by~\cite{Jackson1957a,*Jackson1957b,Herczeg2001}:
\begin{equation}
  A_\beta = \frac{\frac{\rho^2(1\!-\!y^2)}{I\!+\!1} -
    2\rho\sqrt{\frac{I}{I\!+\!1}}(1\!-\!xy)}{(1\!+\!x^2)+\rho^2(1\!+\!y^2)},
  \label{eq:Abeta}
\end{equation}
where $x\approx(\delta-\zeta)/(1-\zeta)$ and $y\approx(\delta+\zeta)/(1+\zeta)$ 
are nonzero in left-right symmetric models.  The SM prediction for 
\isotope[37]{K} \strikethrough{($I\!=\!3/2$)}is found by setting 
$x\!=\!y\!=\!0$ \strikethrough{, viz $A_\beta^\mathrm{SM}=\frac{2}{5}\rho
\left(\rho\!-\!\sqrt{15}\right)/(1\!+\!\rho^2)$}.  With the 
above value of $\rho$ derived from the measured $\mathcal{F}t$ value, 
the result is $A_\beta^\mathrm{SM}=-0.5706(7)$.  The value and sign of $\rho$ 
is such that the 
sensitivity of $A_\beta$ to its uncertainty is reduced compared 
to other observables; e.g., for the $\nu$ asymmetry it is nearly $2\times$ 
bigger, $B_\nu^\mathrm{SM}=-0.7701(18)$.  The value of $\rho$ varies 
considerably among \isotope[37]{K} and the other well-studied 
mirror nuclei (\isotope[19]{Ne}, \isotope[21]{Na} and \isotope[35]{Ar}) 
making each nucleus complementary 
to the others as each will have different dependencies on beyond the SM 
physics.

Recoil-order and radiative corrections to $A_{\beta}$~\cite{Holstein1974,
  *Holstein1974err} are included in our analysis. For isobaric analog 
decays, the induced 1st-order tensor form factor is very 
small (only present because of isospin symmetry breaking), and all but the 
very small induced pseudoscalar and $q^2$ expansion of the Fermi and 
Gamow-Teller form factors~\cite{Towner} are given by the conserved vector 
current (CVC) hypothesis using measured electromagnetic 
moments~\cite{Holstein1974,*Holstein1974err}.  These corrections combine to 
add $\approx-0.0028E_\beta/E_0$ to the expression for $A_{\beta}$.

\begin{figure}\centering
  \includegraphics[width=\columnwidth]{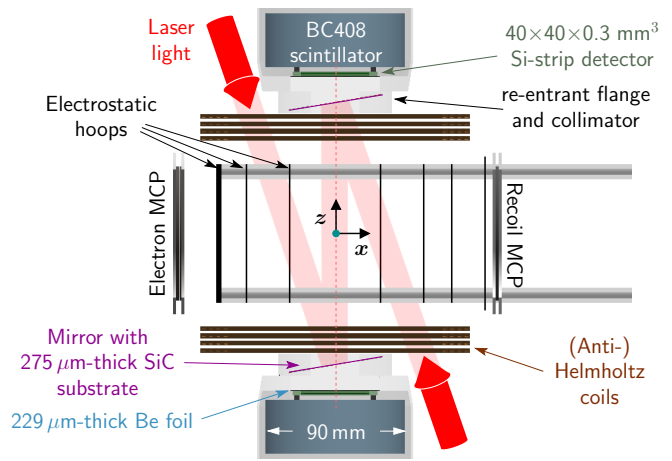}
  \caption{The \trinat~detection
    chamber. To polarize the atoms along the 
    $\beta$-detection ($\hat{z}$-) axis, optical pumping 
    light is brought in at a $\SI{19}{\degree}$ angle with 
    respect to the $\hat{z}$ axis and reflected off thin mirrors 
    mounted within a $\beta$ collimator on the front face of the reentrant 
    flanges.  Thin Be foils behind the mirrors separate the 
    \isotope{Si} strip and 
    scintillator $\beta$ detectors from the \SI{1e-9}{Torr} vacuum of the 
    chamber.   Magnetic field coils provide the Helmholtz (optical pumping,  
    \SI{2}{Gauss}) and anti-Helmholtz (MOT) fields.  Glassy carbon and 
    titanium electrostatic hoops produce a uniform electric field of 150 to 
    \SI{535}{V\per\centi\meter} in the $\hat{x}$ direction to guide shakeoff 
    electrons and ions towards microchannel plate detectors.
    \label{fig:chamber}
  }
\end{figure}

The experiment described here was performed with the \triumf{} Neutral
Atom Trap (\trinat)~\cite{Behr2014HI,Behr2009}.  \strikethrough{\isac, the 
radioactive ion beam facility at \triumf,}
\triumf's radioactive ion beam facility, \isac, 
delivered $8\times10^7$ \isotope[37]{K} ions/s, 
0.1\% of which were neutralized and trapped.  Background from the decay of 
untrapped atoms in the collection MOT was avoided by pushing the 
trapped atoms every second by a pulsed laser beam to a second 
MOT~\cite{Swanson1998} where the precision measurement took place, depicted 
in Fig.~\ref{fig:chamber}.

Once the atoms are collected in the second MOT, we apply a sub-Doppler 
cooling scheme unique to \strikethrough{\isotope{K} atoms} 
potassium~\cite{Landini2011}.  Since the 
atoms can only be polarized while the MOT is off, we alternate between 
periods of trapping and polarizing the atoms.  To optimize the shutoff 
time of the MOT's magnetic field, we employ an alternating-current MOT 
(ac MOT)~\cite{Harvey2008}.  Once atoms are pushed from the first trap 
and cooled, a series of 100 cycles begins, where each cycle consists of
$\SI{1.9}{\milli\second}$ of polarizing the \isotope[37]{K} nuclei and 
collecting polarized decay data, followed by $\SI{3.0}{\milli\second}$ of 
re-collecting the atoms with the ac MOT. \ \ This cycle is repeated with 
the polarization direction ($\sigma^\pm$) \strikethrough{being} flipped 
every \SI{16}{\second}.

While the MOT light and magnetic fields are off, we optically pump the atoms 
on the $D_1$ ($4s_{1/2}\!\rightarrow\!4p_{1/2}$) transition with circularly 
polarized light.  
This technique directly polarizes the nucleus via the hyperfine 
coupling of the atomic and nuclear spins.  
\strikethrough{We measure the degree}
It also lets us measure $P_\mathrm{nucl}$ nondestructively 
by probing the atoms with a pulsed 
\SI{355}{\nano\meter} UV laser and detecting the resulting photoions with 
the recoil MCP detector. The UV photons can only ionize atoms from the $4p$ 
excited state which fully polarized atoms cannot populate, so the rate 
of photoions is a sensitive probe of $P_\mathrm{nucl}$.\ \  
\strikethrough{The result, described in Ref.~\cite{Fenker-NJP-2016},} 
Since $1-P_\mathrm{nucl}$ is small, its determination to 10\% precision is 
sufficient to achieve~\cite{Fenker-NJP-2016,thisPRLsupplmat}: 
${P}_\mathrm{nucl}^{\,\sigma^+}=\SI{99.13\pm0.08}{\percent}$ and 
${P}_\mathrm{nucl}^{\,\sigma^-}=-\SI{99.12\pm0.09}{\percent}$. 
\phantom{\cite*{lonergan,rester}}


The time of flight (TOF) between the photoions and the UV laser pulse 
images the trap along $\hat{x}$, while a delay-line anode readout of the 
MCP provides position sensitivity to image the other axes.  Since the MOT's 
cycling transition produces a relatively large fraction of atoms in the $4p$ 
state, the position of the atoms is well known while the MOT is on.  When 
the MOT light is off, \strikethrough{there are} very few atoms are 
available to be photoionized, 
and the trap position must be inferred from observations immediately before 
and after the polarized phase. From these measurements, we observed that 
\strikethrough{during} the atom cloud moved \SI{0.37\pm0.05}{\milli\meter} 
while expanding from a volume of \SI{2.67\pm0.08}{\milli\meter\cubed} to 
\SI{16.9\pm0.3}{\milli\meter\cubed}. The entire cloud was illuminated by 
the OP light of \SI{20}{\milli\meter} diameter ($1/e^2$) throughout the 
optical-pumping cycle.

To identify decays that occurred within the region of optical pumping, we 
detect the low-energy shakeoff electrons (SOE) by sweeping them with an 
electric field towards an MCP and observing them in coincidence with the 
$\beta^+$. At least one SOE is present for every $\beta^+$ 
decay~\cite{Gorelov2000,Gorelov2005authors} because the \isotope{Ar}$^-$ 
ion is unstable.

To detect the nuclear decay products, we employ a pair of $\beta$ telescopes 
along the vertical polarization axis (Fig.~\ref{fig:chamber}).  Each consists 
of a thin double-sided \isotope{Si}-strip detector (DSSSD) backed 
by a \SI[number-unit-product=\text{-}]{35}{\milli\meter} thick BC408 
scintillator.  The \SI[number-unit-product=\text{-}]{300}{\micro\meter} thick 
\strikethrough{\isotope{Si}-strip detector is} DSSSDs are segmented into 
\SI[number-unit-product=\text{-}]{1}{\milli\meter} strips on both sides, 
\strikethrough{and provides} providing position and $\Delta E$ 
information. Because of its low efficiency 
for detecting $\gamma$ rays, it also suppresses the background from
\SI[number-unit-product=\text{-}]{511}{\kilo\electronvolt} annihilation 
radiation. 

\begin{figure}\centering
  \includegraphics[width=\columnwidth]{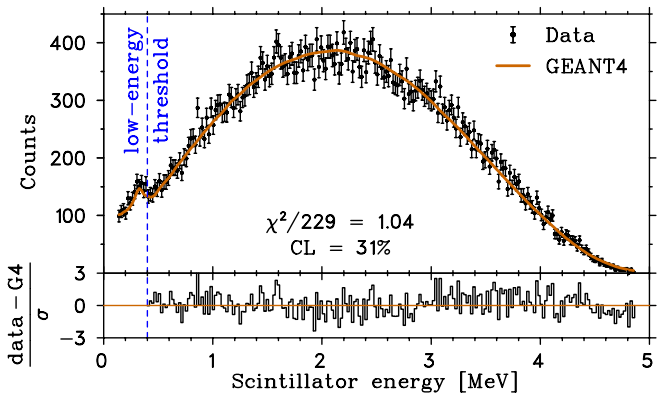}
  \caption{Scintillator spectrum in coincidence with its 
    \strikethrough{Si-strip detector} DSSSD and the electron MCP, 
    showing a very clean selection 
    of $\beta$-decay events originating from the trapping region. The 
    \geant{} comparison shows residuals consistent with statistics. The 
    vertical dashed blue line shows the energy threshold used to exclude 
    Compton-scattered annihilation radiation.\label{fig:scintillator_spectra}
  }
\end{figure}

The plastic scintillators and \strikethrough{silicon-strip detectors} 
DSSSDs were calibrated by comparing the observed spectra to a 
\geant{} simulation.  \strikethrough{In the case of} For the plastic 
scintillators, we assumed a linear calibration and a detector 
resolution with a $1/\sqrt{E}$ dependence.  The calibration was 
\strikethrough{done to} performed using 
the scintillator spectrum in coincidence with a SOE \strikethrough{but} 
without \strikethrough{using} 
adding the energy deposited in
the \strikethrough{silicon-strip detector}
DSSSD\@.  The calibration spectrum included both 
$\beta^+$ events and the Compton edge of the 511-keV annihilation radiation.  
The resulting spectra including the \strikethrough{strip-detector} 
DSSSD coincidence, shown in 
Fig.~\ref{fig:scintillator_spectra} for one detector, agree well with the 
simulation over the entire observed $E_{\beta}$ range.

The asymmetry is calculated by comparing the observed rate of $\beta$ 
particles in the two detectors.  Since the experiment uses two symmetric 
detectors and reverses the sign of the polarization, we use the superratio 
technique which reduces many systematic uncertainties (see 
Refs.~\cite{Gay1992,Plaster2012} for details).

The data analysis was performed blind by temporarily culling an unknown 
fraction, up to $1\%$, of $\beta$-decay events from the analysis. All 
analysis cuts, corrections, and uncertainties were finalized on the 
biased data.  The complete data set was then reanalyzed in this 
predefined way to obtain the final results presented here. 

A detailed representation of the geometry of Fig.~\ref{fig:chamber} was 
included in the \geant{} simulation~\cite{Agostinelli2003,FenkerPhD2016}. The position 
of each decay was randomly sampled from the observed distribution, modeled 
as a Gaussian ellipsoid and included the effects of the cloud's expansion 
and drift.  We used the \texttt{emstandard\_opt3} variation of the standard 
physics lists as well as nondefault values of \SI{1}{\micro\meter} for the 
cut-for-secondaries parameter and  a range factor of $f_R=\num{0.002}$ in 
order to simulate the low-$E_{\beta}$ scattering of $\beta^+$ more 
accurately~\cite{Soti2013}.  The multiple scattering (MSC) of $e^\pm$ was 
simulated with the Urban MSC model of Ref.~\cite{Lewis1950} to avoid the 
nonphysical behavior of the Goudsmit-Saunderson MSC model~\cite{Goudsmit1940} 
observed in Ref.~\cite{Soti2013}.

The simulation was tested by directly comparing the fraction of $\beta^+$ 
that backscattered 
\strikethrough{off} 
out 
of the plastic scintillator.  A large fraction of 
these events have the distinct signature of depositing energy in two different 
pixels of the \strikethrough{silicon-strip detector} DSSSD\@.  The 
number of these backscattered 
events, normalized by the number of events leaving energy only in one pixel, 
was found to 
\strikethrough{agree within 10\% of the measured values}
differ by only $(2.6\pm1.3)\%$ from the measured values~\cite{thisPRLsupplmat}.

\begin{figure}\centering
  \includegraphics[width=\columnwidth]{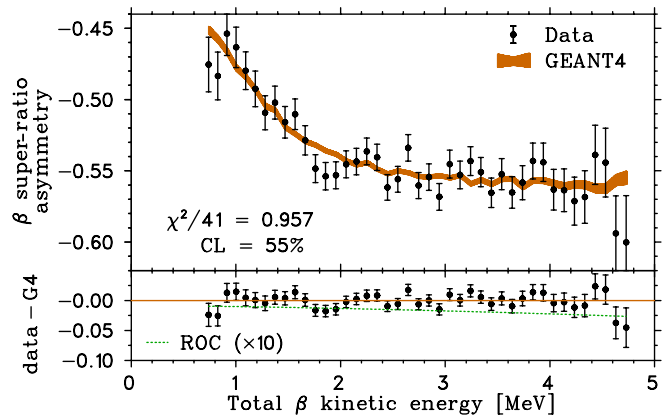}
  \caption{Top: The physics superratio of a subset of the data 
    (points) fit to a \geant{} simulation (filled band, with the width 
    indicating its statistical uncertainty) where the only free parameter 
    was the value of $\rho$. Bottom: Difference between the data and \geant, 
    and the small size of the recoil-order+radiative corrections (ROC).
    \label{fig:asymmetry_overlay_D}
  }
\end{figure}

Events are considered in the asymmetry analysis if they (i) occur during 
the portion of the duty cycle that the atoms are fully polarized, (ii) 
have a valid \strikethrough{strip-detector} DSSSD hit as well as energy 
deposited in  the 
scintillator, and (iii) are in coincidence with a SOE.\ \ The four spectra 
for upper (lower) detector and spin up (down) are compared at a number of 
energy 
bins using the superratio technique to calculate the observed asymmetry 
shown in Fig.~\ref{fig:asymmetry_overlay_D}.  The energy dependence is 
dominated by \strikethrough{mostly from} the $\beta$'s finite 
helicity [$p_\beta/E_\beta$ of Eq.~\eqref{eq:kin}]. 
The observed asymmetry is compared to the \geant{} simulation in order to 
obtain the best-fit results for the input asymmetry.

Although our geometry is very open, $\beta$ scattering off of volumes such 
as the opposite $\beta$ telescope, electrostatic hoops, etc.\ (see 
Fig.~\ref{fig:chamber}), must be accounted for by \geant.  Simulations 
indicate that 1.60\% of accepted events scattered by $\geq\!24^\circ$ before 
being detected, leading to an effective 
$\langle\cos\theta\rangle=0.9775$~\cite{thisPRLsupplmat,FenkerPhD2016}.  
The \geant{} simulations therefore apply a 2.30\% correction due to $\beta$ 
scattering.  
Using a combination of our data and some from the literature, we assign a 
systematic uncertainty which is 5.6\% of the correction (see 
Table~\ref{tab:abeta_systematics}), 
as explained in the Supplemental Material~\cite{thisPRLsupplmat}. 


Accounting for our measured
$\langle P\rangle=99.13(9)\%$~\cite{Fenker-NJP-2016},
a simultaneous fit to all of our data yields a best-fit value 
$A_\mathrm{obs}=-0.5699(13)$ with $\chi^2/123=0.82$.  
\strikethrough{This statistical 
uncertainty does not include the background correction which we discuss next.}

\begin{figure}\centering
  \includegraphics[width=\columnwidth]{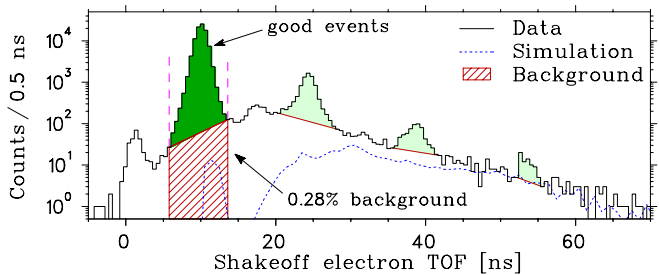}
  \caption{Shakeoff electron TOF spectrum with respect to 
    the $\beta^+$, showing all data at an electric field of 
    \SI{150}{V\per\centi\meter}.  
    This spectrum constrains the production of metastable Ar$^-$ with 
    $\tau=\SI{260(25)}{\nano\second}$~\cite{ben-itzhak-PRA38-1988} 
    to be less than 4\%, while the TOF cut eliminates any possible contribution.
    Overlaid is a simulation (dotted line) of 
    the TOF from atoms that escaped the trap before decaying from an 
      electrostatic hoop, where the 
      only free parameter is the normalization fixed to times 
      $\geq43~\mathrm{ns}$.  While this 
    \strikethrough{Considering 
    this is not a fit, the} simulation reproduces the longer TOF very 
    well, \strikethrough{although} it does not explain all of the background 
    (red hatched area) under the main peak of good events within our TOF 
      cuts (dashed vertical lines).
    \label{fig:soe_cuts_2d1d}
  }
\end{figure}

The TOF spectrum of SOEs with respect to the $\beta^+$ 
(Fig.~\ref{fig:soe_cuts_2d1d}) has the expected large, narrow peak near 
$t=10~\mathrm{ns}$, the good events we use in our analysis. The peaks 
at $24, 39$, and $53~\mathrm{ns}$ come from 
electrons that do not fire the MCP, but produce a secondary $e^-$ that is 
re-collected by the electric field which is registered by the MCP.\ \ We 
can simulate most of the broad TOF structure to be background from decays 
of atoms stuck to the SiC mirrors and electrostatic hoops.  The same 
simulation suggests an unresolved peak at $12~\mathrm{ns}$ from the 
electrode nearest the trapping region, but this does not \strikethrough{fully} 
account for the majority of the total background 
\strikethrough{fraction of $0.28\%$} under the good peak: 0.28\%.  
We conservatively assume \strikethrough{conservatively} that this 
unknown background is either fully polarized or unpolarized atoms 
and make a correction 
$A_{\beta}=A_\mathrm{obs}\times1.0014(14)$.

\begin{table}\centering
  \caption{Uncertainty budget for $A_{\beta}$.  Each entry is given as the 
    absolute 
    uncertainty, and correction factors and the range varied are 
    listed where applicable.
    Polarization uncertainties, detailed in Ref.~\cite{Fenker-NJP-2016}, 
    are statistically independent. 
    \label{tab:abeta_systematics}
  }
  \begin{tabular}{lll@{}l@{}l@{}l@{}l@{}l@{}l@{}l@{}l}
    \toprule
    \multicolumn{6}{l}{Source \hspace*{12em}} & \multicolumn{2}{c}{Correction} &  \multicolumn{3}{c}{Uncertainty} \\
    \midrule
    \multicolumn{11}{l}{Systematics} \\
   ~& \multicolumn{5}{l}{Background} &~~~~~&1.0014 &~~~&~~~& 0.0008 \\[0.25em]
    & \multicolumn{5}{l}{$\beta$ scattering\tsups{a}} && 1.0230 & &&  0.0007\\[0.25em]
    &&&&& \multicolumn{3}{l}{position (typ $\lesssim\!\pm20\,\mu\mathrm{m}$)} & && 0.0004 \\
    & \multicolumn{4}{l}{Trap {\footnotesize($\sigma$\textsuperscript{+}\hspace*{0.05em}vs\,$\sigma$\textsuperscript{--})}\,\opara{3}{0}{0}{2em}}& 
    \multicolumn{5}{l}{sail velocity (typ $\lesssim\!\pm30\,\mu\mathrm{m\!/\!ms}$)} & 0.0005 \\
    &&&&& \multicolumn{3}{l}{temperature (typ $\lesssim\!\pm0.2\,\mathrm{mK}$)} & && 0.0001 \\[0.25em]
    &&& \multicolumn{5}{l}{radius\tsups{a} ($15.5^{+3.5}_{-5.5}~\mathrm{mm}$)} & &&  0.0004 \\
    & \multicolumn{2}{l}{Si-strip \opara{3}{0}{0}{2em}}& 
    \multicolumn{5}{l}{energy agreement ($\pm3\sigma\rightarrow\pm5\sigma$)} & &&  0.0002 \\
    &&& \multicolumn{5}{l}{threshold ($60\rightarrow 40~\mathrm{keV}$)} & && 0.0001\\[0.25em]
    & \multicolumn{9}{l}{Shakeoff electron TOF region ($\pm3.8\rightarrow\pm4.6\,\mathrm{ns}$)} & 0.0003\\[0.25em]
    &&&& \multicolumn{4}{l}{SiC mirror\tsups{a} ($\pm6~\mu\mathrm{m}$)} & && 0.0001\\
    &\multicolumn{3}{l}{Thicknesses} \opara{3}{0}{0}{2em}& \multicolumn{4}{l}{Be window\tsups{a} ($\pm23~\mu\mathrm{m}$)} & && 0.000\,09\\
    &&&& \multicolumn{4}{l}{Si-strip\tsups{a} ($\pm5~\mu\mathrm{m}$)} & && 0.000\,01\\[0.25em]
    & \multicolumn{9}{l}{Scintillator only vs.\ $E+\Delta E$\tsups{a}} &  0.0001 \\
    & \multicolumn{9}{l}{Scintillator threshold ($400\rightarrow 1000~\mathrm{keV}$)} &  0.000\,03 \\
    & \multicolumn{9}{l}{Scintillator calibration ($\pm0.4~\mathrm{ch/keV}$)} &  0.000\,01 \\[0.5em]
    \multicolumn{8}{l}{Total systematics}  &&  \multicolumn{2}{l}{0.0013}\\[0.25em]
    \multicolumn{8}{l}{Statistics}         &&  \multicolumn{2}{l}{0.0013} \\[0.25em]
    \multicolumn{8}{l}{Polarization}       1.0088 &&  \multicolumn{2}{l}{0.0005} \\[0.25em]
    \midrule
    \multicolumn{8}{l}{Total}  1.0338 && \multicolumn{2}{l}{0.0019} \\
    \bottomrule
    \multicolumn{11}{l}{\tsups{a}Denotes sources that are related to 
      $\beta^+$ scattering.}
  \end{tabular}
\end{table}

Although the superratio technique greatly reduces the systematic 
uncertainties (e.g.\ the cloud position, $\beta$ detector differences, and 
$\beta$ scattering), this cancellation is not exact.  Independently, we 
adjusted the trap position, size, temperature, drift velocity, and other 
parameters within the \geant{} simulation, obtaining the systematic 
uncertainties shown in Table~\ref{tab:abeta_systematics}.  

The final result is 
\begin{equation}
  A_\beta = -0.5707\left(13\right)_{\mathrm{syst}}\left(13
  \right)_{\mathrm{stat}}\left(5\right)_{\mathrm{pol}},
\end{equation}
where the third uncertainty combines the systematic and statistical 
uncertainties on the polarization measurement~\cite{Fenker-NJP-2016}. 
This result has the lowest relative uncertainty of any measurement of the 
$\beta$ asymmetry in a nuclear system to date.  Since the simulation 
includes the recoil-order and radiative corrections, this result may be 
directly compared to $A_\beta^\mathrm{SM}$ given earlier.

\begin{figure}\centering
  \includegraphics[width=\columnwidth]{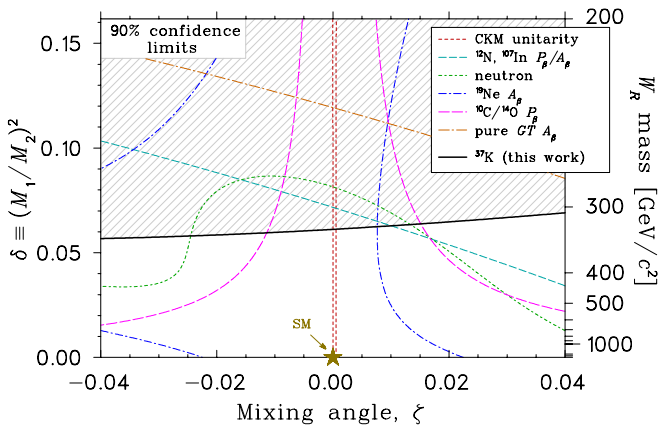}
  \caption{Constraints on manifest L-R symmetric models 
    from nuclear and neutron~\cite{PDG2016} $\beta$ decay: CKM 
    unitarity~\cite{Hardy2015}; the ratio of $\beta^+$ polarization to 
    $A_\beta$ of \isotope[12]{N} and \isotope[107]{In}~\cite{Thomas2001,
      SeverijnsNaviliat2011}; $A_{\beta}$ of mixed GT/F 
    \isotope[19]{Ne}~\cite{Severijns2008,Calaprice1975,Broussard2014,
      Naviliat-Cuncic2009}; the $\beta^+$ polarization of \isotope[10]{C} 
    compared to \isotope[14]{O}~\cite{Carnoy1990}; and the weighted average 
    of $A_\beta$ from three recent pure-GT 
    cases~\cite{Wauters2009,Wauters2010authors,Soti2014}.\label{fig:RHC}
  }
\end{figure}

Figure~\ref{fig:RHC} shows the allowed parameter space in the manifest 
left-right model. We vary $\rho$ at each $(\zeta,\delta)$ coordinate 
to minimize the $\chi^2$ over all observables ($\mathcal{F}t$, $A_{\beta}$ 
and $B_{\nu}$). The \isotope[37]{K} limit includes our previous $B_\nu$ 
measurement~\cite{Melconian2007}, but is dominated by the present $A_\beta$ 
result.

Assuming $\zeta\!=\!0$ from other experiments (particularly 
Ref.~\cite{Hardy2015}), our result implies $\delta=0.004^{+45}_{\ -4}$ 
and a mass for a $W_R$ coupling to right-handed $\nu^R$ greater than 
\SI{340}{\giga\electronvolt/c^2} at 90\% confidence, 
\strikethrough{an} 
a slight 
improvement over the $P_\beta/A_\beta$ 
\SI{310}{\giga\electronvolt/c^2} limit~\cite{Thomas2001,SeverijnsNaviliat2011}. 
Much of the parameter space in left-right symmetric models has been 
excluded by other measurements.  
Constraints from polarized muon decay~\cite{Bueno2011} are relaxed if the 
$\nu_\mu^R$ is heavy (as e.g.\ in Ref.~\cite{Asaka2005}). 
LHC searches directly exclude $W_R$ with 
mass $<3.7~\mathrm{TeV}/c^2$ if the right-handed gauge coupling 
$g_R\!=\!g_L$~\cite{PDG2016}, while our \isotope[37]{K} results imply 
\strikethrough{$g_R<7.6$} 
$g_R<8$ 
for a \SI{4}{\tera\electronvolt/c^2} $W_R$.  Manifest models with 
$M_{W^\prime}\!<\!M_W$ and $V_\mathrm{ud}^R$ considerably less than unity are 
also constrained by $\beta$ decay correlations~\cite{Thomas2001}.

\begin{figure}\centering
  \includegraphics[width=\columnwidth]{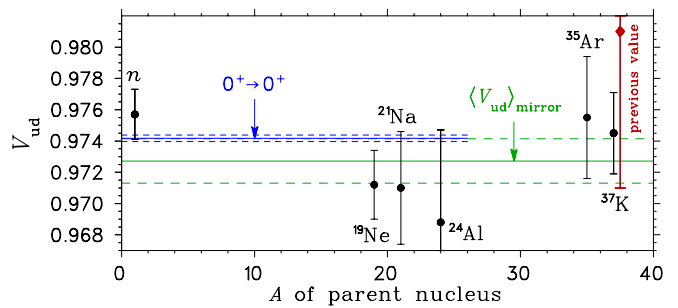}
  \caption{Measurements of $V_\mathrm{ud}$ comparing the 
    values from the neutron~\cite{PDG2016}, 
    \isotope[24]{Al}~\cite{Adelberger1985}, 
    and the $T\!=\!1/2$ mirror nuclei: 
    \isotope[19]{Ne}~\cite{Broussard2014}, 
    \isotope[21]{Na}~\cite{Grinyer2015}, 
    \isotope[35]{Ar}~\cite{Naviliat-Cuncic2009}, the previous value for 
    \isotope[37]{K}~\cite{Shidling2014}, and the present work.  The 
      averages (uncertainties) in $V_\mathrm{ud}$ determined from 
      \zerotozero~\cite{Hardy2015} and mirror transitions are shown as 
      the solid (dashed) lines.\label{fig:Vud}
  }
\end{figure}

If we make the assumption that the SM completely describes the $\beta$ 
decay of \isotope[37]{K}, we can use the result to test the CVC hypothesis.  
Combining the present result for $A_\beta$ with the previous measurement of 
$B_\nu$~\cite{Melconian2007}, we find 
$\rho=0.576(6)$.  
This, in combination with the $\mathcal{F}t$ value of 
Ref.~\cite{Shidling2014}, leads to
$V_\mathrm{ud}=0.9744(26)$  
for \isotope[37]{K}, a \strikethrough{$4.2\times$} greater than $4\times$ 
improvement over the previous value~\cite{Shidling2014}. Isospin-mixing 
calculations~\cite{Severijns2008} contribute $0.0004$ to this uncertainty, 
which only grows to $0.0005$ if the span between the isospin-tuned shell 
model of Ref.~\cite{Severijns2008} and the density functional of 
Ref.~\cite{Konieczka2016} is taken as the uncertainty.  We compare this 
determination of $V_\mathrm{ud}$ to other nuclear $\beta$-decay measurements 
in Fig.~\ref{fig:Vud}.  Our \isotope[37]{K} result has the same accuracy as 
\isotope[19]{Ne}~\cite{Broussard2014} and improves a CVC test at 
$I\!>\!1/2$~\cite{Adelberger1985}.  Combining the four values from the 
$T\!=\!1/2$ mirror transitions leads to a new average 
$\langle V_\mathrm{ud}\rangle_\mathrm{mirror}=0.9727(14)$, 
only \strikethrough{$6.6\times$} $6.7\times$ 
less precise than the \zerotozero{} result~\cite{Hardy2015} and slightly 
better than the neutron.

We have used a highly polarized, laser-cooled source of \isotope[37]{K} 
to measure the $\beta$ asymmetry in its decay to be $A_\beta=-0.5707\pm0.0019$, 
placing limits on the mass of a hypothetical $W_R$ coupling to right-handed 
$\nu$'s as well as improving the value of $V_\mathrm{ud}$ from mirror 
transitions.  The high precision 
of our nuclear polarization measurement on the \strikethrough{\itshape in situ} 
atom cloud is 
enabling a further program of improved $A_{\beta}$, $B_{\nu}$, and recoil 
asymmetry measurements. 

\begin{acknowledgments}
We acknowledge \triumf/\isac{} staff, in particular for TiC target 
preparation, and the remaining authors of Ref.~\cite{Fenker-NJP-2016} 
for previous polarization development.  Supported by the Natural Sciences 
and Engineering Research Council of Canada, the Israel Science Foundation, 
and the U.S. Department of Energy, Office of Science, Office of Nuclear 
Physics under Award No.\ DE-FG03-93ER40773 and No.\ DE-FG02-11ER41747.  
\triumf{} receives federal funding via a contribution agreement through 
the National Research Council of Canada.
\end{acknowledgments}


%
%

\newpage\ 
\newpage
\renewcommand\thefigure{S\arabic{figure}}
\setcounter{figure}{0}    
\renewcommand{\abstractname}{Abstractorama}

\begin{minipage}{\columnwidth}\centering
  {\bfseries\boldmath Supplemental Material for {\itshape Precision Measurement of the 
      $\beta$ Asymmetry in Spin-Polarized $^{37}$\!K Decay}}\\
  B. Fenker, A. Gorelov, D. Melconian, J.A. Behr, M. Anholm, D. Ashery, 
  R.S. Behling, I. Cohen, I. Craiciu, G. Gwinner, J. McNeil, M. Mehlman, 
  K. Olchanski, P.D. Shidling, S. Smale, and C.L. Warner
\end{minipage}

\section{Atomic Physics}
Given the precision measurement of our Letter, we include here some
additional details about the atomic physics methods for the interested
reader. Certain atomic effects produce negligible
uncertainties on the determination of the
nuclear polarization $P$ needed to deduce
the value of $A_{\beta}$ for $^{37}$K,
and we explore these through more detailed
measurements made possible by larger quantities of stable $^{41}$K atoms.
We take the opportunity to provide
some qualitative guides to our detailed publication on our polarization 
methods in Ref.~\cite{blah1}.

There are several features of our optical pumping and probing method that 
we want to emphasize. We probe the small unpolarized fraction, so not much 
precision is required. Our probe is parasitic --- unlike most more direct 
methods, it does not alter the polarization during probing. We also measure 
$P$ throughout the duty cycle of polarization, so we can choose the best times 
of the duty cycle to determine $A_{\beta}$. 

\subsection{Optical pumping tests on stable \tsups{41}K}
\subsubsection{Qualitative description}
For optical pumping of small densities of atoms, there are two depolarization 
mechanisms. We measure optically the degree of imperfectly circularly polarized 
light (see Section 2.3 of Ref.~\cite{blah1}).  We then fit the excited state 
population mentioned in our Letter for a parameter $B_x$, an average magnetic 
field perpendicular to the optical pumping $\hat{z}$ axis. 
The result, after a full detailed characterization of optical pumping using 
the well-established optical Bloch equations (OBE, described in 
Ref.~\cite{blah1}), determines the population of unpolarized atomic states. 
Most important is the population of two almost-pumped ground states 
($F\!=\!2~M_F\!=\!1$ and $F\!=\!1~M_F\!=\!1$) with nuclear polarization $1/2$ 
and $5/6$: determining their population supplies the precision needed for 
this Letter. Larmor precession governed by $B_x$ 
does not change $F$, while the measured imperfect circular polarized light 
can change $F$ by optical pumping, so once we quantify the two depolarization 
mechanisms the OBE's give us the populations we need. 

The tail/peak ratio of the excited state population determines $1\!-\!P$.
Given that $1\!-\!P$ is less than $0.01$, and that the nuclear polarization is
$\geq0.5$ for the two almost-pumped unpolarized states, we
only need on order 10\% accuracy on the tail/peak ratio to achieve the result of
Ref.~\cite{blah1},  $1\!-\!P = 0.0087\pm0.0009$. If there were no excited state
fluorescence in $^{41}$K at long times in Fig.~\ref{fig:S1}
(or photoions in $^{37}$K in Ref.~\cite{blah1}),
the polarization would be 100\% with no uncertainty. 


Thus effectively, a single parameter, $B_x$, is fit to the $^{37}$K excited
state population data.
All other parameters are fixed by independent measurements on $^{37}$K,
and high-statistics independent data on $^{41}$K in the same geometry.
The $^{41}$K data we describe in this Supplemental Material is helpful in
lending confidence to our model, but in the
end not essential to the $^{37}$K polarization result.

\subsubsection{Time dependence of $B_x$}
We mention in Section 2.3 of Ref.~\cite{blah1} that we measure the time-dependence 
of $B_x$ with Hall probes as the MOT quadrupole $B$ field falls. This provides 
reasonable accuracy, albeit with vacuum system open without detectors installed, 
and suggests the depolarization from this component is unimportant in the 
part of the duty cycle used for $A_{\beta}$ data.

\begin{figure}
    \includegraphics[angle=90,width=1.0\linewidth]{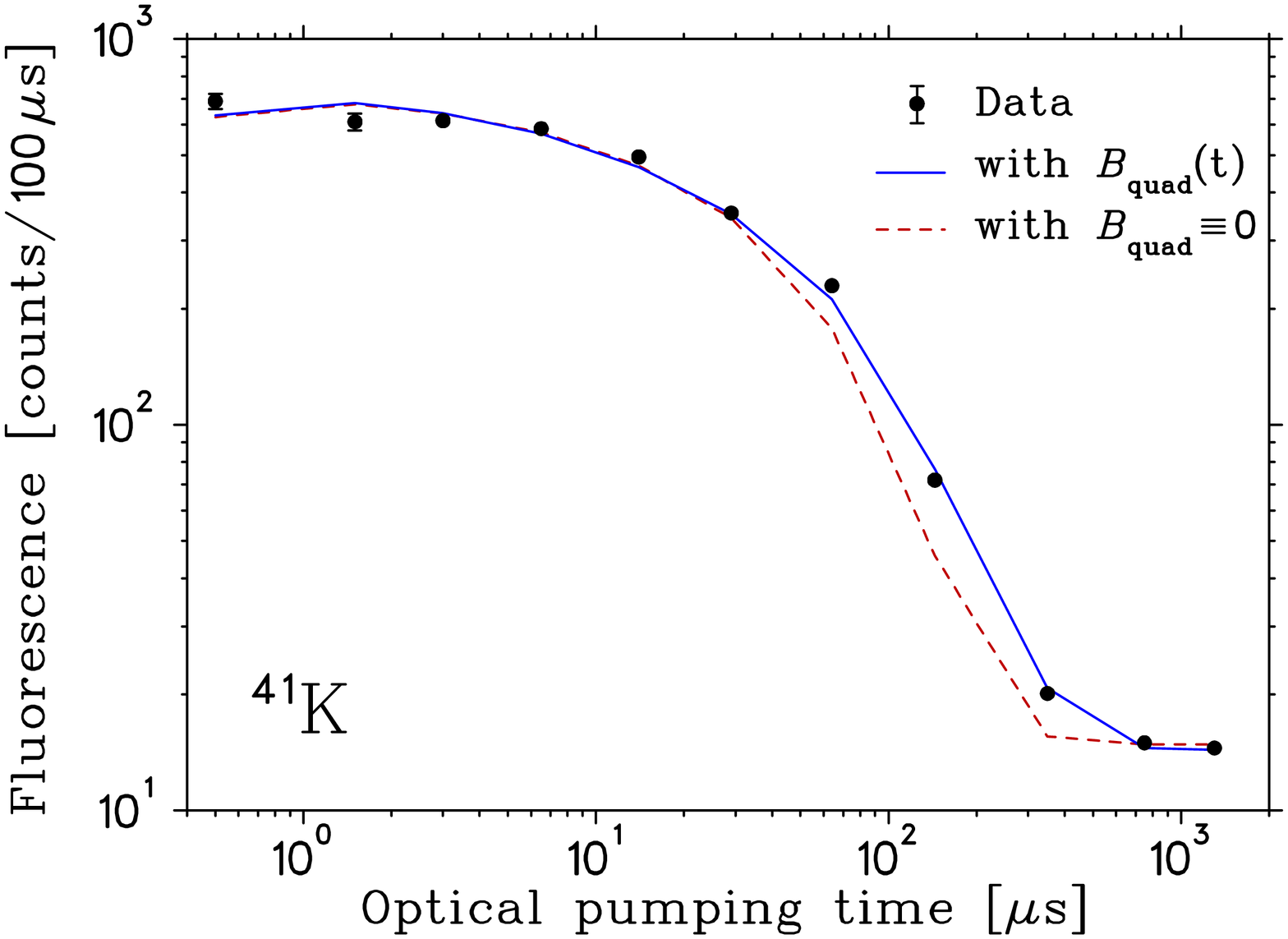}\\
\caption{Fluorescence of $^{41}$K in $4S_{1/2}$ to $4P_{1/2}$ transition during 
  optical pumping. Note the log-log scale showing the peak fluorescence, the 
  region dominated by falling $B_x[t]$, and the tail due to imperfect 
  polarization.\label{fig:S1}}
\end{figure}

To test this with all detectors in place, we optically pump stable $^{41}$K 
atoms. $^{41}$K has almost the same hyperfine structure as $^{37}$K, so after 
adjusting laser frequencies experimentally the OBE predict almost the same 
results. We show in Fig.~\ref{fig:S1} the dependence of the fluorescence of the 
$4P_{1/2}$ state as a function of time, along with optical pumping 
calculations including the time dependence of $B_x$.  The region from 60 to 
$200~\mu$s after the optical pumping starts is better modelled if we include 
this time-changing $B_x$, with the fall time fixed to the Hall probe 
measurement of $\tau\!=\!130~\mu\mathrm{s}$. In Fig.~\ref{fig:S1} the MOT 
quadrupole field was turned off at $-250~\mu\mathrm{s}$. We waited longer 
for the MOT field to decay away before we started optical pumping $^{37}$K 
(see Fig.~10 of Ref.~\cite{blah1}), and then simply waited for this field 
to decay away to take the $A_{\beta}$ data.  The tail/peak ratio, and hence 
the deduced polarization of $^{37}$K at OP times used for $\beta$ decay, do 
not depend on whether or not the decaying MOT quadrupole field is included 
in the theory. Nor is the goodness of fit with and without this effect 
changed in the $^{37}$K photoion data (Fig.~8 and Fig.~10 of 
Ref.~\cite{blah1}).

\begin{figure}
  \includegraphics[angle=90,width=1.0\linewidth]{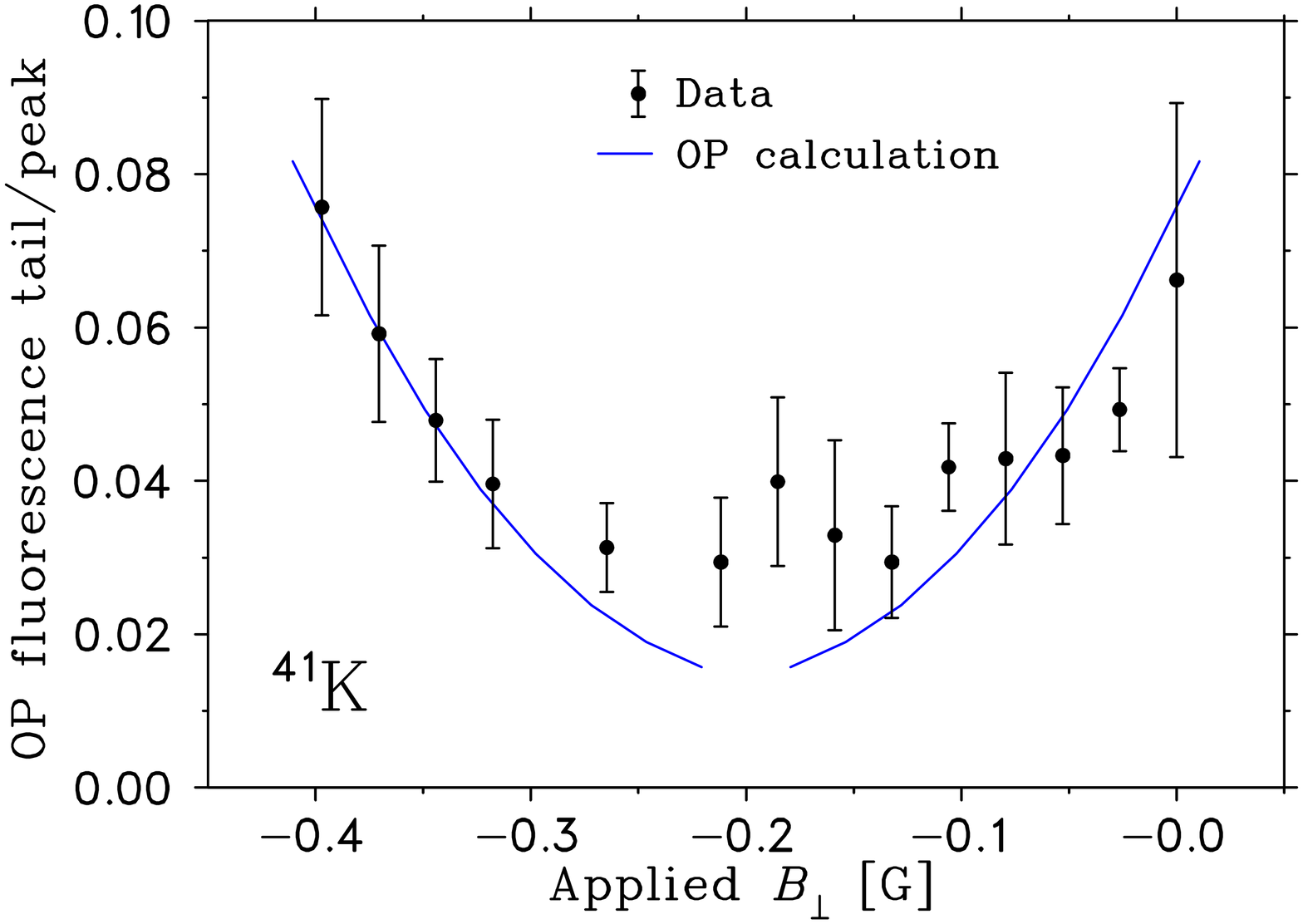}\\
  \caption{Optical pumping tail/peak ratio for $^{41}$K, optimized by changing 
    one set of uniform-field Helmholtz coils.\label{fig:S2}}
\end{figure}

With data of this sort, we can tune parameters to optimize the polarization.
Figure~\ref{fig:S2} shows optimization of one perpendicular uniform magnetic 
$B$ field by trimming the current through a Helmholtz coil along one 
perpendicular axis, after similar optimization of the other axis.  
This effectively aligns the total $\vec{B}$ field with the OP laser light 
axis. The OP laser light axis is in turn aligned mechanically with the 
$\beta$ detector axis, to optical alignment accuracy of $1~\mathrm{mm}$ in 
$1~\mathrm{m}$, or $0.001~\mathrm{rad}$, producing negligible misalignment 
accuracy on $\cos\theta_{\beta \hat{I}}$ of less than $10^{-6}$.

We also learn that the polarization difference from unity depends 
quadratically on applied $B_\perp$, in agreement with our OP calculation. 
The result in Fig.~\ref{fig:S2} is consistent with a small average 
horizontal field 
that is not zeroed out with our uniform applied field. In our $^{37}$K data 
we are also able to fit for this effective $B_x$, and find it is consistent 
with the values found for $^{41}$K, as would be expected since the atom clouds 
are located at almost the same position in the apparatus.

\paragraph{Spatial gradients of the B field}
Given the dying remnants of the time-changing MOT quadrupole field,
it is natural to consider whether spatial gradients of the magnetic field
can make gradients of the polarization across the atom cloud. In particular,
a finite $dP/dz$ could in principle perturb $A_{\beta}$ significantly. However, 
the possible residual $dB/dz$ of $<0.01~\mathrm{G/cm}$ detunes the optical 
pumping laser by negligible Zeeman shifts, so negligible $dP/dz$ is produced. 
For measurement of future $\beta$-decay observables, polarization
gradients along the other axes are being studied (by fast CMOS camera) and
minimized (by the standard trick of unbalancing Helmholtz coils).

\subsection{Metastable Ar$^-$ atoms}
If nothing else happens, $\beta^+$ decay of a potassium atom populates a
negative Ar ion. The ground state of this ion is, of course, unstable, and
dissociates in negligible time. There is a known metastable state of the Ar$^-$ 
ion with lifetime $\tau\!=\!260~\mathrm{ns}$~\cite{blah2}. 
We mention here that this state makes negligible contribution to $A_{\beta}$ 
systematic uncertainties.

The angular distribution of the $\beta^+$ is quite different in singles versus
in coincidence with the recoil (i.e.\ the $\nu$). So it is important
in measuring $A_{\beta}$ that the shakeoff electrons be detected with no bias
from the recoil direction. A metastable Ar$^{-}$ could in principle move
in $z$ first before releasing the electron, thus biasing the critical
coincidence.

We can fit for a tail with the known lifetime in the $\beta$--shakeoff electron
TOF spectra like Fig.~4 of our Letter (but to longer times than shown).
The population of the Ar$^{-1}$ is less than 4\%, which could produce a less 
than 0.08\% correction to $A_{\beta}$ using a $40$-$\mathrm{mm}$ diameter 
shakeoff electron detector.  However, such a tail is excluded by the time 
width of the $\beta$-shakeoff coincidence in Fig.~4 used for $A_{\beta}$, so 
the possible distortion to $A_{\beta}$ vanishes.

\section{\boldmath$\beta$ (back)scattering}
A primary concern of any $\beta$ asymmetry measurement is the effect of 
$\beta$ scattering before entering the detector.  These events will have 
an apparent initial direction that is incorrect and will therefore 
bias the results -- especially in the case of large-angle backscatters.  A 
separate 
publication is in preparation where we will describe in greater detail our 
estimates of these effects~\cite{blah3}, but in order for 
the reader of our Letter to understand how we arrived at the correction and 
uncertainty for $\beta$ scattering in Table I of the Letter, we provide the 
plots comparing our data to our \geant{} simulation which led to these results. 

One comparison of the data to \geant{} could be made by looking at events 
where the $\beta$ backscatters off of one double-sided Si-strip detector 
(DSSSD) into the both the scintillator and DSSSD of the opposite $\beta$ 
telescope. However, given the small ($\sim0.25\%$) solid angle for a $\beta$ 
to go from one telescope to the other, these events are extremely rare, 
$\lesssim10^{-4}$ of non-scattering events.  The effect of $\beta$ 
backscattering on the $A_\beta$ measurement in our geometry is highly 
suppressed:  the 20 candidate events in our data set are too few to serve 
as a meaningful benchmark for \geant.   Although negligible, these events 
were vetoed in the analysis of $A_\beta$. 



A much more frequent type of backscatter we are able to measure  experimentally 
are events in one $\beta$ telescope where the scintillator and two pixels 
in the corresponding DSSSD are all above threshold~\cite{blah4}.  
These ``scintillator-backscatter'' events correspond to a $\beta$ entering 
a pixel in the DSSSD of one of the $\beta$ telescopes, leaving energy in 
the plastic scintillator, and then backscattering out through a different 
pixel of the same DSSSD detector.  
Note that this is a very clean measurement: the triple-coincidence between 
the shake-off electron MCP, the DSSSD and the scintillator greatly suppresses 
$\gamma$ events and other backgrounds, and in particular the shake-off electron 
coincidence ensures the decay occurred from the trap.

\begin{figure}\centering
  \includegraphics[angle=90,width=\linewidth]{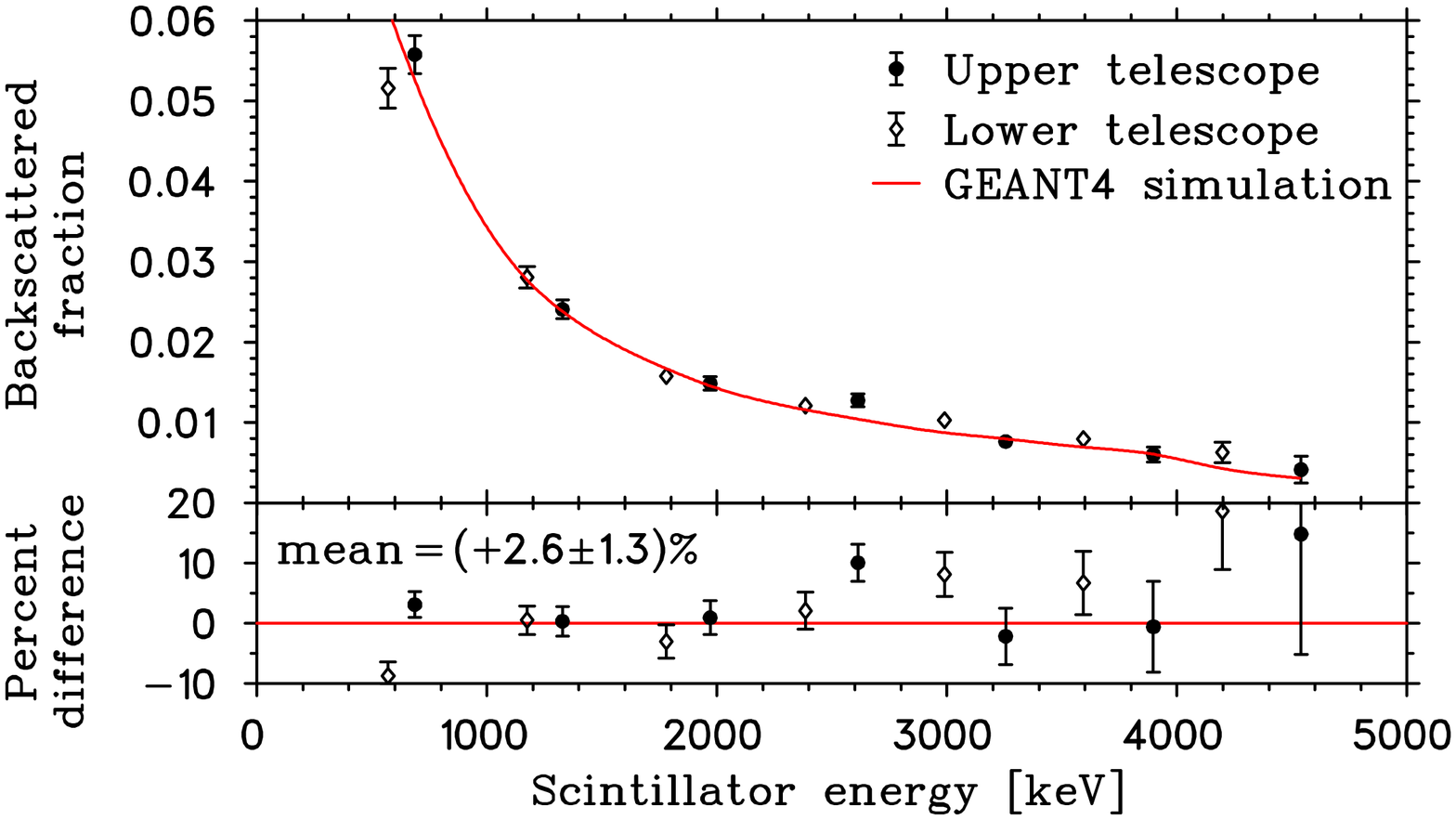}\\
  \caption{Comparison of \geant{} with the observed fraction of events 
    backscattering out of the scintillator through a 2nd pixel in the 
    DSSSD $\Delta E$ detector.  The bottom plot shows the percent difference 
    between \geant{} and observations.\label{fig:S3}}
\end{figure}

Figure~\ref{fig:S3} shows the fraction of scintillator-backscatter 
events normalized to the number of good events as observed by each 
$\beta$ telescope.  These are compared to the fraction predicted by the 
\geant{} simulation, which can be seen to be quite favourable when using the 
non-standard \geant{} options listed in the Letter: the average difference 
is only $(+2.6\pm1.3)\%$ over the range of energies considered in our 
analysis.  This unique measurement of backscattering out of plastic 
scintillator serves as our benchmark for testing the efficacy of our 
\geant{} simulations.  

The same \geant{} simulation is used to predict the effect of $\beta$ 
scattering on the $A_\beta$ measurement.  Given the position of an event in 
the DSSSD and assuming the decay occurred from the trap center, we are able to 
calculate the angle between the polarization direction and the momentum of 
the $\beta$.  If the $\beta$ scattered before entering the detector, this 
calculated angle will be wrong, most dramatically for events which 
backscattered off of a volume opposite the telescope in which it was 
detected.  To estimate the effect, we performed a simulation looking at 
events which fired the $\beta$ telescope in the same direction as the 
initial nuclear polarization (so $\cos\theta_\mathrm{calc}\approx1$) and 
compared this to the actual $\cos\theta$ of the generated event.  

\begin{figure}\centering
  \includegraphics[angle=90,width=\linewidth]{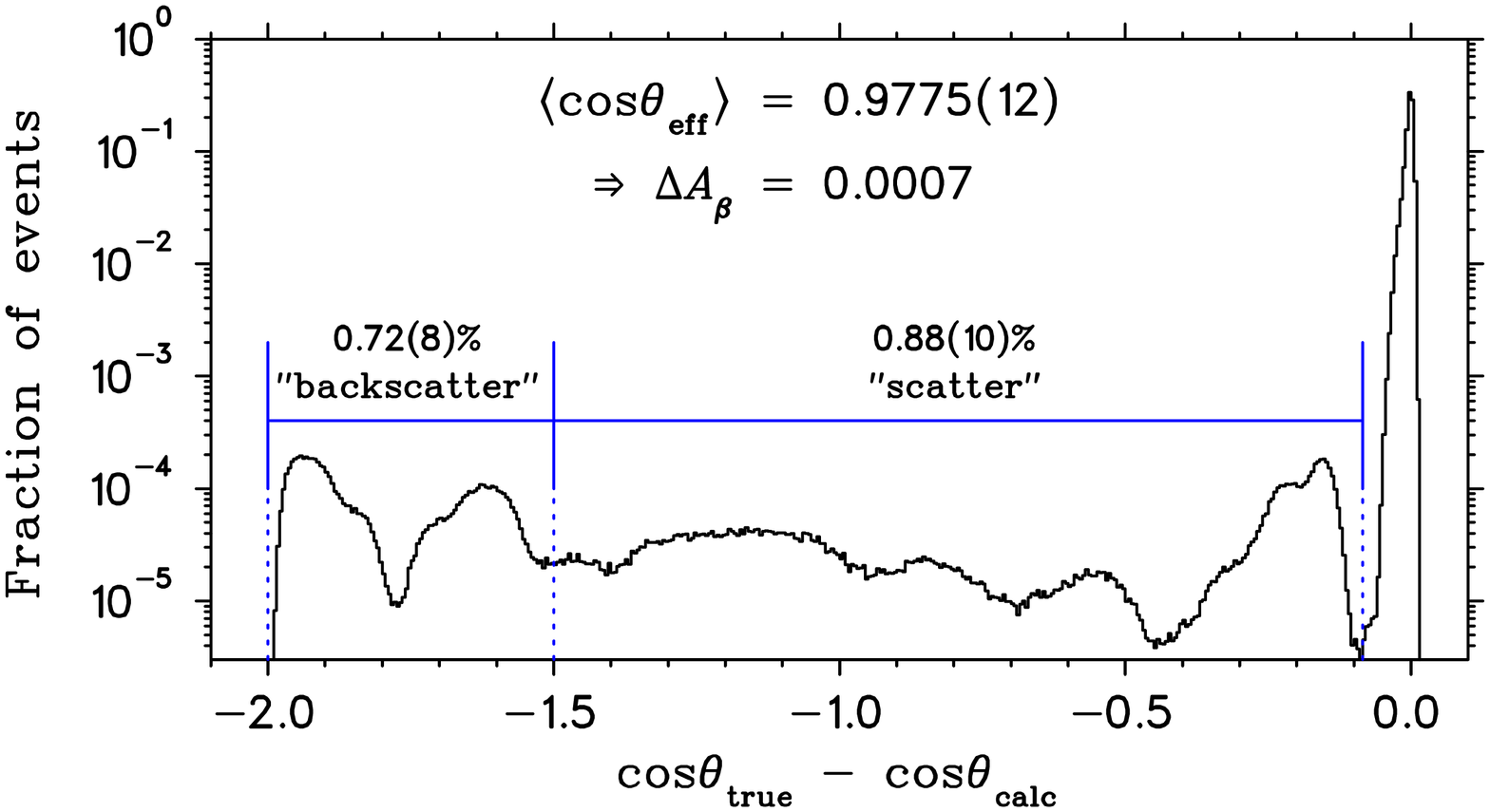}\\
  \caption{\geant{} simulation showing the 
    effect on the $A_\beta$ measurement due to $\beta$ scattering.  
    The dominant peak at $\Delta\cos\theta\approx0$ 
    are events which entered the $\beta$ telescope directly from the trap; the 
    events below this peak are ones where the $\beta$ scattered before 
    entering the detector and which lead to an incorrect angle reconstruction.
    We have divided these events into two regions: 0.72(8)\% of events we 
    labelled ``backscatter'' events, and 0.88(10)\% ``scatter'' (see text).  
    Instead of the true $\cos\theta$, $\beta$-scattering effects lead to an 
    effective $\cos\theta$ that is attenuated by 2.3\%.
    \label{fig:S4}}
\end{figure}

Figure~\ref{fig:S4} shows the distribution of simulated events as a 
function of the true $\cos\theta$ minus that which we calculate based on the 
position in the DSSSD\@. The main peak at $\Delta\cos\theta\approx0$, 
containing $98.40(12)\%$ of the spectrum, are events which entered the $\beta$ 
telescope with minimal scattering; most of the width of this peak is due 
to the finite position resolution of the DSSSD ($1~\mathrm{mm}$ strips) 
and finite size of the cloud of atoms.  The 
events below this main peak correspond to events which scattered off the 
opposite $\beta$ telescope, one of the electrostatic hoops and/or one of the 
other volumes shown in Fig.~1 of the Letter.  

All together, \geant{} predicts that scattered events reduce the observed 
asymmetry by $1/\langle\cos\theta_\mathrm{eff}\rangle=1.0230$.  Our analysis, 
based on these \geant{} simulations, includes this 2.30\% correction for 
$\beta$ scattering.  
To assign a systematic uncertainty, we consider three 
regions in Fig.~\ref{fig:S4}:  ``not scattered'' events are those where 
$\Delta\cos\theta\geq-0.085$; ``backscattered'' events are those where 
$\Delta\cos\theta\leq-1.5$; and the rest, $-1.5<\Delta\cos\theta<-0.085$ are 
``scattered''.  We varied the fraction of events in the ``scattered'' and 
``backscattered'' regions to estimate a systematic uncertainty on 
$\langle\cos\theta_\mathrm{eff}\rangle$.  For the ``backscattered'' region, 
we use our result from Fig.~\ref{fig:S3} to assign an 
uncertainty 
of $5.1\%$, the $2\sigma$ upper-limit of the difference shown.  We have no data 
of our own to constrain the fraction of ``scattered'' events, so for these we 
assign a 
$10\%$ uncertainty, consistent with the accuracy of a \geant{} simulation we 
ran compared to literature data on few MeV electron transmission through 
thin materials into angles of  
$10-75~\mathrm{degs}$~\cite{blah5,blah6}.  The result is a 
$\pm0.0012$ uncertainty on $\langle\cos\theta_\mathrm{eff}\rangle$ and an 
absolute systematic uncertainty of $\pm0.0007$ on $A_\beta$, which is 
$5.6\%$ of the total correction.  This is the 
systematic uncertainty assigned directly to $\beta$ scattering in Table~I 
of the Letter.  Note that there are five other entries in this table of 
uncertainties (labelled with a superscript ``a'') which also contribute 
to $\beta$ scattering, albeit to a lesser extent and less directly.




%

\end{document}